\documentclass[aps,prb,twocolumn,oneside,floatfix,sort&compress,
superscriptaddress,showpacs,nobalancelastpage]{revtex4-1}

\usepackage{graphicx,xcolor}
\usepackage[utf8]{inputenc}
\usepackage{amssymb,bbm}
\usepackage[reqno]{amsmath}
\usepackage{bbm}

\usepackage[normalem]{ulem}

\begin{document}
\title{ Lanczos algorithm with Matrix Product States for dynamical correlation functions}

\author{P.E. Dargel}
\affiliation{Institut für Theoretische Physik, Georg-August-Universität Göttingen, 37077 Göttingen, Germany}
\author{A. W\"ollert}
\affiliation{Physics Department, Arnold Sommerfeld Center for Theoretical Physics, and Center for NanoScience, Ludwig-Maximilians-Universit\"at M\"unchen, D-80333 M\"unchen, Germany}
\author{A. Honecker}
\affiliation{Institut für Theoretische Physik, Georg-August-Universität Göttingen, 37077 Göttingen, Germany}
\affiliation{Fakultät für Mathematik und Informatik, Georg-August-Universität Göttingen, 37073 Göttingen, Germany}
\author{I.P. McCulloch}
\affiliation{Centre for Engineered Quantum Systems, School of Mathematics and Physics, The University of Queensland, St Lucia QLD 4072, Australia}
\author{U. Schollw\"ock}
\affiliation{Physics Department, Arnold Sommerfeld Center for Theoretical Physics, and Center for NanoScience, Ludwig-Maximilians-Universit\"at M\"unchen, D-80333 M\"unchen, Germany}
\author{T. Pruschke}
\affiliation{Institut für Theoretische Physik, Georg-August-Universität Göttingen, 37077 Göttingen, Germany}
                
\date{\today}                 

\begin{abstract}
The density-matrix renormalization group (DMRG) algorithm can be adapted to the
calculation of dynamical correlation functions in various ways which all 
represent compromises between computational efficiency and physical accuracy.
In this paper we reconsider the oldest approach based on a suitable 
Lanczos-generated approximate basis and implement it using matrix product 
states (MPS) for the representation of the basis states. The direct use 
of matrix product states combined with an ex-post reorthogonalization method 
allows to avoid several shortcomings of the original
approach, namely the multi-targeting and the approximate representation of the 
Hamiltonian inherent in earlier Lanczos-method implementations in the DMRG 
framework, and to deal with the ghost problem of Lanczos methods, 
leading to a much better convergence of the spectral weights and 
poles. We present results for the dynamic spin structure factor of the spin-1/2
antiferromagnetic Heisenberg chain. A comparison to Bethe ansatz results in 
the thermodynamic limit reveals that the MPS-based Lanczos approach is much more
accurate than earlier approaches at minor additional numerical cost.
\end{abstract}
\maketitle

\section{Introduction}
{\em Dynamical correlation functions.} A central question in physics  is the calculation of time-dependent 
correlation functions of the type
${\rm i} G_A(t',t) \equiv \langle 0 | A^\dagger(t') A(t) | 0 \rangle$, 
where $A$ is some operator, $|0\rangle$ the ground state  and 
$t' \geq t$ for causality. In the case of a system given by a time-independent Hamiltonian $H$, 
it is often more convenient to Fourier transform to frequency space, arriving 
at some dynamical correlation function $G_A(\omega+{\rm i}\eta)$, taking 
the form
\begin{equation}
G_A(\omega+{\rm i}\eta) = \langle 0 | A^\dagger 
\frac{1}{\epsilon_0+\omega+{\rm i}\eta-H} A|0\rangle ,
\end{equation}
where $\epsilon_0$ is the ground state energy and $\eta$ an infinitesimal positive 
number introduced for convergence. The precise calculation of such and similar 
dynamical correlation functions is highly important in condensed matter 
physics as they can be directly compared to experiments measuring the response 
of a many-body system to an external perturbation, for example in 
angle-resolved photoemission spectroscopy or inelastic neutron scattering, upon
suitable choices of $A$. From a methodological point of view, dynamical 
correlation functions are important ingredients for several approximation 
schemes in condensed matter theory, for example cluster perturbation
theory (CPT), variational cluster approximation (VCA), dynamical mean field
theory (DMFT) and others (see Ref.~\onlinecite{Potthoff2012} for a 
comprehensive review). In these approaches, one needs so-called impurity or 
cluster solvers, which provide results for Green's functions entering the 
further many-body calculations of the method. Usually, analytical solutions 
are not available and one has to resort to numerical approaches; dynamical 
correlation functions have been calculated in numerical methods as diverse as 
the numerical renormalization group, exact diagonalization, quantum Monte Carlo 
and the density-matrix renormalization group (DMRG). The latter is at the 
basis of this work.

{\em Dynamical correlation functions in DMRG.} 
The density-matrix renormalization group (DMRG) \cite{White1992,White1993,Schollwock2005,Schollwoeck2011} has established itself as the most powerful method 
for the calculation of ground states of strongly correlated systems in one
dimension, such that its extension to dynamical quantities such as 
$G_A(\omega+{\rm i}\eta)$ has been an obvious focus of research. However, the 
calculation of dynamical quantities within the DMRG is still a difficult task, 
for which various approaches have been developed which show different 
combinations of advantages and disadvantages, which we will briefly outline.
 
Hallberg\cite{Hallberg1995} made a first attempt to tackle this task by using 
a continued fraction expansion of dynamical correlation functions based on a 
Lanczos algorithm which had already been successfully employed in the context 
of exact diagonalization calculations.\cite{Haydock1972,Gagliano1987} This algorithm is 
fast, easily implemented in a ground state DMRG program, and resolves 
correlation functions $G_A(\omega+{\rm i}\eta)$ whose spectral weight is 
concentrated in relatively sharp $\omega$-peaks very well. Its drawback is that
it does not resolve correlators with broadly distributed spectral weight very 
well. However, it has found numerous successful applications.\cite{Nishiyama1999,Zhang1999,Yu2000,Okunishi2001,Garcia2004}  Recently, some of the authors have proposed an adaptive Lanczos method\cite{Dargel2011} which improves over the original method of Hallberg. 

Shortly after Hallberg's original proposal, Ramasesha {\em et al.} \cite{Ramasesha1996} and later  
K\"uhner and White \cite{Kuhner1999} introduced the correction vector method,
which since then has successfully been applied to many model 
systems.\cite{Pati1999,Kuhner2000,Jeckelmann2002,Jeckelmann2003,Benthien2004,Nishimoto2004,Raas2004,Raas2005,Friedrich2007,Benthien2007,Smerat2009,Weichselbaum2009,Peters2011,Grete2011} The correction vector method achieves high precision, albeit at much higher numerical cost than the Lanczos method. Fourier-transforming time-dependent DMRG data offers another approach to dynamical spectral functions.\cite{White2004,Sirker2005,Garcia2006,Pereira2006,Pereira2008,Pereira2009,Barthel2009,Bouillot2011,Karrasch2011}  However, accessing long time scales, to obtain good frequency resolution, is limited by a rapid increase of entanglement and the resulting increasing demand of computational resources. Very recently, dynamical spectral functions were also calculated using an expansion of spectral functions in terms of Chebychev polynomials.\cite{Holzner2011} While early results look very promising, the full potential of this method is still unexplored.
 
{\em DMRG in the MPS framework.} In an independent development, it has been recognized early in the history of 
DMRG that DMRG produces quantum states that take the very specific form of 
so-called matrix product states (MPS).\cite{Ostlund1995} Indeed, DMRG -- with minor variations to the original formulation -- can be recognized to be simply a variational search for the lowest energy state within that state class.\cite{Dukelsky1998,Verstraete2004} In the context of ground state DMRG this insight allows a more elegant formulation of the method and highlights connections to entanglement theory, but does not yield a more performing algorithm.\cite{Schollwoeck2011} This story changes, however, once one is confronted with an algorithmic situation where more than one quantum state of interest is being calculated, not just say the ground state. In the original DMRG, the problem of an acceptable approximation for several states at one time is solved by the procedure of multi-targeting: the reduced density operator that governs the choice of reduced bases in DMRG is set up from a weighted sum of reduced density operators formed for each state individually. This implies that none of the states will be represented with the accuracy the method could have achieved for it alone. In the MPS formulation, each state carries its own implicit (optimal) choice of reduced basis, hence is approximated more precisely; moreover, the formalism automatically deals with the fact that each state has been approximated differently. Another advantage is that upon the introduction of matrix product operators (MPOs) the Hamiltonian can be expressed exactly whereas in DMRG it is expressed in the reduced basis chosen by DMRG to represent the state(s) of interest.\\
{\em Combining Lanczos and MPS.} Given the range of options for calculating dynamical correlation functions in the DMRG framework, it would be highly desirable if the Lanczos method, which is computationally very undemanding and has very few fine-tuning parameters, could be improved in its range of applicability while maintaining low computational cost. The purpose of this paper is to provide a substantial improvement of the current Lanczos-based methods. This will be achieved by making use of the methodological progress due to the use of MPS and MPOs and a few additional modifications:

(i) As already noted in Ref.~\onlinecite{Dargel2011}, the MPS formulation should be much more suited than original DMRG for the Lanczos method, because it generates a long sequence of quantum states that call for optimal representation beyond multi-targeting; moreover, they are obtained by subsequent applications of the Hamiltonian, such that an exact representation of the Hamiltonian should be very useful. In this work we will therefore use the MPS 
formulation\cite{Ostlund1995,Dukelsky1998,Takasaki1999,Verstraete2004,
Verstraete2004a,Vidal2004,McCulloch2007,Schollwoeck2011} of DMRG to implement the Lanczos algorithm. Every Lanczos state will be represented by one MPS and the only approximation comes from the compression of the state. 

(ii) The problem of numerical loss of orthogonality between
Lanczos states is at first sight exacerbated in the
MPS formulation due to the required, but 
inexact compression of MPS after addition of MPSs or application of the Hamiltonian operator to an MPS (both inevitable in the Lanczos algorithm) in comparison
to the exact Lanczos method. It can, however, be removed
by an ex-post reorthogonalization method.

(iii) The method allows direct access to spectral poles and weights without broadening, and we can do finite-size extrapolations to the thermodynamic limit using a special rescaling procedure. 

The new approach combines the advantages of a very good representation of individual states, largely eliminating the so-called ``ghost'' problem of Lanczos approaches, and a removal of the broadening issue.

{\em Outline of the paper.}
In Section \ref{sec:model}, we will introduce the model to be used for the exposition of the new method and as a test case, the spin-$\frac{1}{2}$ Heisenberg model. In Section \ref{sec:Lanczos}, we will review the basic Lanczos algorithm for dynamical correlation functions, and its inherent problems (finite-size broadening, loss of orthogonality) and briefly discuss the status of the current Lanczos-based methods, before we introduce the new approach (Section \ref{sec:newLanczos}). In order to test the performance of the algorithm we have calculated the dynamical structure factor of the antiferromagnetic spin-$\frac{1}{2}$ Heisenberg chain (Section \ref{sec:results}). This quantity is not characterized by a single magnon band, where the pole structure is very simple, but rather a multi-spinon continuum. Such complex structures are generally hard to access by the Lanczos method, and hence provide a relevant test case to see how this approach performs for systems with non-trivial dynamical properties. To this end we compare our results with those from the thermodynamic Bethe ansatz. We find that the MPS-based approach leads to much more accurate results than earlier Lanczos methods. The paper is concluded by a discussion (Section \ref{sec:discussion}).

\section{Model}\label{sec:model}
As model to test our algorithm we choose the antiferromagnetic spin-$\frac12$ Heisenberg chain with
Hamiltonian
\begin{align}
H=&\sum_i \textbf{S}_i \textbf{S}_{i+1}\notag\,. 
\end{align}
Here $\textbf{S}_i$ denote the usual spin-operators at site $i$ and we have set the antiferromagnetic exchange coupling 
constant $J=1$ in this work. The model has a $SU(2)$ symmetry, which has been exploited in the DMRG 
program.\cite{McCulloch2002,McCulloch2007} We are interested in calculating the dynamical spin structure factor
\begin{align}
S(k,\omega)&=I_{\textbf{S}^{\phantom{\dagger}}_k}(\omega)\notag\\
&=-\frac{1}{\pi}\lim_{\eta\rightarrow 0}\text{Im}\langle 0|\textbf{S}^{\dagger}_k\frac{1}{\omega-H+\rm{i}\eta+\epsilon_0}\textbf{S}^{\phantom{\dagger}}_k|0\rangle\notag\\
&=\frac{1}{\pi} \sum_n |\langle E_n|\textbf{S}_k|0 \rangle|^2 \delta(\omega -\epsilon_0- \epsilon_n)\;,\notag 
\end{align}
where $|0\rangle$ denotes the groundstate with energy $\epsilon_0$. If we use open boundaries (obc) in the DMRG, we have to define the spin-operator $\textbf{S}_k$ in Fourier space (in the spirit of Ref.~\onlinecite{Benthien2004}) as 
\begin{align}
\textbf{S}_k= \sqrt{\frac{2}{L+1}}\sum^L_{j=1}\sin(jk)\textbf{S}_j
\end{align}
with quasi-momentum $k=l\pi/(L+1)$, $l\in\mathbb{N}\mod L$. In this paper we also present results for periodic boundary conditions (pbc), even though the DMRG is not that efficient for pbc. The spin-operator is then given by 
\begin{align}
\textbf{S}_k= \sqrt{\frac{2}{L}}\sum^L_{j=1}\exp({\rm i}jk)\textbf{S}_j
\end{align}
with momentum $k=l\pi/L$, $l\in\mathbb{Z}\mod L$.
The asymptotics of the dynamical structure factor are known from the analytical solutions,\cite{Cloizeaux1962,Faddeev1981,Muller1979,Karbach1997,Caux2006} and can be summarized as follows: 
\begin{itemize}
\item the dominant part of the structure factor comes from the two-spinon contributions. These are located
in an interval bounded by
\begin{align}
\omega_1 = \frac{\pi}{2}|\sin k| \text{ and } \omega_2=\pi|\sin \frac{k}{2}| \label{bounds}
\end{align}
\item the dynamical structure factor is known\cite{Karbach1997,Caux2006} to diverge as:
\begin{equation}
S(k,\omega)\propto(\omega-\omega_1)^{-\frac{1}{2}}\sqrt{\ln(1/(\omega-\omega_1))}\label{div_nopi}
\end{equation}
for $k\neq \pi$, respectively
\begin{equation}
S(\pi,\omega)\propto \omega^{-1}\sqrt{\ln(1/\omega)}\label{div}
\end{equation}
for $k=\pi$.
\end{itemize}

\section{Current Lanczos algorithms}
\label{sec:Lanczos}

\subsection{The basic Lanczos algorithm}
\label{subsec:basicLanczos}

We are interested in calculating dynamic correlation functions at $T=0$. In the Lehmann representation they can be expressed via
\begin{align}
I_A(\omega)&=-\frac{1}{\pi}\lim_{\eta\rightarrow 0}\text{Im}\langle 0 | A^\dagger\frac{1}{\omega-H+\rm{i}\eta+\epsilon_0}A\phantom{\dagger}|0\rangle\label{dyncorr} \\
&= \frac{1}{\pi} \sum_n |\langle E_n|A|0 \rangle|^2 \delta(\omega -\epsilon_0- \epsilon_n) \,.\notag 
\end{align}
Here $|E_k\rangle$ denote the eigenfunctions of  the Hamiltonian $H$ with eigenenergy $\epsilon_k$ 
(spectral poles).
$A$ is the operator that defines the correlation function. For small systems one can obtain the full set of eigenvectors and eigenenergies by exact diagonalization. However,  this approach is limited rather quickly
by the exponential growth of the Hilbert space.  In order to circumvent this problem one can use an iterative 
eigensolver like for example the Lanczos algorithm. Within the Lanczos algorithm one recursively generates
an orthogonal basis that spans the Krylov space. The recursion formula\footnote{Within the application of the method to Green's functions the recursion formula is sometimes called Lanczos-Haydock recursion formula. See Refs.~\onlinecite{Haydock1972,Haydock1980} for further details. } for the so-called Lanczos vectors is
given by
\begin{align}
|f_{n+1}\rangle&=H|f_n\rangle-a_n|f_n\rangle-b^2_n|f_{n-1}\rangle
\notag\\
a_n&=\langle f_n|H|f_n\rangle/\langle f_n|f_n\rangle, \notag\\
b^2_n&=\langle f_n|f_n\rangle/\langle f_{n-1}|f_{n-1}\rangle,\qquad
b_0=0 \, .\label{recursion}
\end{align}
 The full Hamiltonian is then mapped onto this Krylov space resulting in a  tridiagonal effective Hamiltonian.
 \begin{align}
 H_{\text{eff}}=\begin{pmatrix}
  a_0 & b_1 & 0 & \dots & 0\\
  b_1 & a_2 & b_2 & \ddots  & \vdots \\
  0 & b_2 & \ddots & \ddots & 0\\
  \vdots & \ddots & \ddots & \ddots & b_{n-1}\\
  0 & \dots & 0 &  b_{n-1} & a_{n-1}
 \end{pmatrix}  \,. 
 \end{align}
  
Because of the tridiagonal form this Hamiltonian can be easily diagonalized and it can be shown that the 
extremal eigenvalues are good approximants\cite{Bai2000} to the extremal eigenvalues of the original system.

This procedure, originally developed for finding the extreme eigenvalues of large sparse Hamiltonians matrices, can now be adapted to the calculation of dynamical correlation functions.\cite{Gagliano1987} For such functions as in (Eq.~\eqref{dyncorr}), we are actually not interested
in the full set of eigenvectors but only those vectors with non-zero overlap with the vector $A|0\rangle$. 
Therefore one starts the Lanczos algorithm with the vector $|f_0\rangle = A|0\rangle/\langle 0| \hat{A}^\dagger \hat{A} | 0 \rangle$, so that only eigenvectors and energies are created that give a non-zero contribution to the dynamical correlation function. If one diagonalizes $H_{\text{eff}}$ obtained in this way, its eigenvectors and eigenvalues give direct access to the (lowest) poles and spectral weights of Eq.~{\eqref{dyncorr}. In order to obtain a  ``smooth'' spectral function one can then broaden these delta functions.  Alternatively, one can also calculate directly a 'smooth' spectral function by using the continued fraction expansion
\begin{equation}
    I_{A}(\omega) = -\frac{1}{\pi}\lim_{\eta\rightarrow 0}\text{Im} \frac{\langle 0| \hat{A}^\dagger \hat{A} | 0 \rangle}{
    z-a_{0}-\frac{b_{1}^2}{z-a_{1}-\frac{b_{2}^2}{z-\ldots}}} ,
    \label{eq:continuedfraction}
\end{equation}
with $z=\epsilon_{0}+\omega+{\rm i}\eta$. Both methods to obtain a 'smooth' spectral function are equivalent if one uses a Lorentz broadening of the delta peaks.
The Lanczos method was used for our test case, the antiferromagnetic spin-$\frac{1}{2}$ Heisenberg chain (with periodic boundary conditions) by Fledderjohann {\em et al.} \cite{Fledderjohann1995} and later by Karbach {\em et al.}\cite{Karbach1997} to obtain spectral weights and poles for chains up to $28$ sites. 

How well does this work? If applied to exact Hamiltonians and states, the method is obviously limited by system size, as all ``exact diagonalization'' methods are. But there are more subtle issues.

(i) The Lanczos algorithm will give the best convergence to the extremal eigenvectors that are contained in the starting vector. The interior eigenvectors will only converge after a large number of Lanczos iterations. This is in most cases not possible as numerical errors will destroy 
the orthogonality of the Lanczos vectors. This is the well-known ``ghost problem''\cite{Bai2000,Cullum1985} which leads to spurious (double) eigenvalues. Reorthogonalization and restarting methods are usually applied to tackle this problem.

(ii) We are mainly interested in dynamical correlation functions in a continuous form (in the thermodynamic limit). In order to get a continuous form from our finite size discrete dynamical correlation functions we can use the two above described methods: Continued fraction expansion or Lorentzian broadening of the spectral poles. For large systems and large broadenings this should give a decent approximation of the broadened correlation function in the thermodynamic limit. Extracting the $\eta\rightarrow 0$ limit a posteriori by removing the convolution  necessarily requires at first a finite size scaling $L\to \infty$ of the broadened data. Without a finite size scaling one would essentially try to 'deconvolute back' to a finite number of delta peaks. Furthermore the deconvolution is well-known to be a numerically ill-defined problem - there is no unique solution and the best solution can only be given by probability arguments (See Ref.~\onlinecite{Raas2005} for details). In summary the extrapolation to the interesting thermodynamic limit $\eta\to0, L\to \infty$ is far from trivial especially as the two limits are not to be interchanged.}

This discussion is not specific to the Lanczos method, identical small shifts off the real frequency axis are mandatory in the correction vector method, too. In fact, the $\eta$-broadening is also a feature of spectral functions obtained from time-dependent DMRG where the finite time-range of simulations is damped out by an artificial damping factor $e^{-\eta t}$ which leads to Lorentzian broadening in frequency space. In the approach using Chebychev polynomials, damping, while taking a slightly different form, is also inevitable.\cite{Holzner2011} \\
(iii) Knowing the explicit positions of the spectral poles for a finite system is advantageous, because it allows, among other things, to perform proper finite size scaling for low-energy excitations and clearly identify possible gaps in the spectrum. Furthermore, Holzner {\em et al.}\cite{Holzner2011} suggested recently to calculate the spectral weights for
chains of different length and to use a systematic rescaling of the weights to approximate the dynamical 
correlation function in the thermodynamic limit. The idea of this rescaling is to simply translate the definition of
the spectral function reading ``spectral weight per unit frequency interval'' into a mathematical formula.\cite{Holzner2011}
We adapt their idea and use it for the spectral weights extracted with the Lanczos method. We rescale each of the weights $\Omega_n=|\langle E_n|\textbf{S}_k|0 \rangle|^2$ by the width of the energy interval 
\begin{align}
\Delta_n=(\epsilon_{n+1}-\epsilon_{n-1})/2
\end{align}
the pole is located in. For the lower bound $\omega_1$ of the spectral function this definition has to be 
replaced by $\Delta_1=(\epsilon_{2}+\epsilon_{1})/2-\omega_1$.  Using this rescaling scheme all spectral 
weights will now lie on the spectral function calculated in the thermodynamic limit. Combining the spectral weights/poles of chains with different length, one can then generate a high resolution approximation to the spectral function in the thermodynamic limit and by this get a controlled approach to the problematic limit  $\eta\to0, L\to \infty$.

\subsection{Lanczos with standard/adaptive DMRG}
\label{sec:oldLanczos}

Let us briefly discuss how available Lanczos methods in the DMRG framework operate, regarding the broadening and loss of orthogonality issues as well as additional, DMRG specific, approximations.

The idea to use the Lanczos algorithm within standard DMRG  was 
first put forward by Hallberg\cite{Hallberg1995} following the procedure outlined above, with $|0\rangle$ provided by a ground state DMRG calculation; operations $H|\psi\rangle$ are provided in DMRG such that the Lanczos algorithm can be implemented easily. Even if one assumes that $|0\rangle$ is available at precision close to exact diagonalization, which is true to a very good approximation,
an additional approximation of this approach is that in the calculation of the Lanczos vectors (essentially $H^n |f_0\rangle$) $H$ is available only approximately and usually optimized for the operation $H|0\rangle$. DMRG, if applied naively, will therefore introduce increasingly severe and systematic errors for the higher Lanczos vectors. A partial remedy is provided by the multi-targeting procedure, by which one finds a good approximate representation of several Lanczos vectors (which changes also the approximations made in the representation of $H$). The quandary one finds itself in is that if one targets few Lanczos vectors, those are represented quite accurately, but all others quite badly; if one targets many Lanczos vectors, all are represented at similar, but quite low accuracy.  Loss of orthogonality occurs as in every Lanczos procedure, but can be mended by reorthogonalization of the Lanczos vectors; as DMRG keeps them in a common basis representation, this can be done easily, but the price is the low accuracy of each individual Lanczos vector due to multiple targeting. Mostly the spectral function is presented
as a smooth curve with an artificial broadening, but not as the set of poles and weights naturally following from the Lanczos representation. This makes sense as the accuracy of individual pole positions and spectral weights is not so high in view of the systematic errors.

Recently, some of the authors have proposed an adaptive Lanczos method\cite{Dargel2011} within a DMRG framework that is based on the same approach as Ref.~\onlinecite{Hallberg1995}, but calculates the Lanczos vectors adaptively (i.e., using a multi-targeting approach, but restricting the Lanczos vectors to be targetted to the last three ones which occur in the recursion; this implies a continuous change of basis as the algorithm evolves). They could show that this approach is more efficient than the original formulation and allows one to directly calculate the position of the poles respectively the spectral weights for Green's function without any prior broadening because of higher accuracy. It must be stated however, that this approach is potentially vulnerable to the ever-present problem of Lanczos ``ghosts'', the appearance of spurious eigenvalues due to loss of orthonormality between Lanczos vectors. The adaptive basis changes make reorthogonalization schemes very difficult if not effectively impossible.

\section{Lanczos with Matrix Product States (MPS)}
\label{sec:newLanczos}
Choosing the MPS formulation of DMRG in the Lanczos method will completely eliminate the need for multiple targeting, giving additional accuracy, introduce an exact representation of $H$, and regain the possibility of a reorthogonalization scheme, lost in the adaptive Lanczos method. 

The algorithm proceeds exactly as before, the ground state $|0\rangle$ is written in MPS representation (as can be obtained from a standard ground state DMRG with minor modification or in a variational MPS approach).
\begin{align*}
|\psi\rangle = \sum\limits_{\sigma_1..\sigma_n} A[\sigma_1]A[\sigma_2]..A[\sigma_{n-1}]A[\sigma_n]|\sigma_1..\sigma_n\rangle
\end{align*}

 $H$ is written as a matrix product operator (MPO); for examples, see Ref.~\onlinecite{Schollwoeck2011,McCulloch2007}. If we consider the recursion relation, we have to keep in mind 
that both the application of $H$ to a state and the addition of two MPS (as occurs in the Lanczos procedure) lead, if carried out exactly, to a new MPS, but with a dimension that is {\em larger} than that of the original MPS: under addition, dimensions add; under application of an MPO to an MPS, the dimensions multiply (MPO dimensions are often 4 to 6). This is to be seen in contrast to the same operations in the conventional DMRG framework, where these operations do not increase the numbers of DMRG states to be kept (i.e. the MPS matrix dimension), at the price of an approximate representation of the Hamiltonian $H$ which is adapted only to specific Lanczos states and introduces quite severe inaccuracies for Lanczos states where $H$ has been applied often. It is at this point that MPS based Lanczos dynamics has to introduce its only approximation, namely the compression of the large-dimension MPS back to the original dimension at some loss of accuracy. This compression is carried out iteratively until the compressed state has become stable (for standard procedure, see Ref.~\onlinecite{Schollwoeck2011}).  This approximation avoids the systematic errors due to an approximate Hamiltonian and is in practice less severe. Multiple targeting is eliminated.

Within the MPS formulation we have found the best way to calculate the Lanczos states as MPS, but still the approximation due to the compression of the Lanczos states  will lead to the ghost problem because the compression will increase the amount of orthogonality loss. In the MPS approach, this problem is not as bad as in the adaptive DMRG approach,\cite{Dargel2011} but it is still exacerbated in comparison to the standard DMRG Lanczos method by Hallberg. The advantage is that this problem can be cured in comparison to the inexact Hamiltonian representation.  There are many approaches to resolve the loss of orthogonality. The two most popular ones are total or partial re-orthogonalization and restarting procedures.\cite{Bai2000}
In a partial or total re-orthogonalization procedure one tries to re-orthogonalize the current Lanczos state to the previous ones by 
\begin{align}
|\tilde \psi_n\rangle &=|f_n\rangle-\sum^{n-1}_i \langle f_n|\psi_i\rangle |\psi_i\rangle, \quad 
|\psi_n\rangle &= \frac{1}{N}|\tilde \psi_n\rangle \label{reortho}
\end{align}
after several steps or after each step.  Here, $N=(\langle\tilde\psi_n|\tilde\psi_n\rangle)^{1/2}$ accounts for
the proper normalization. We have tried partial and total re-orthogonalization (not shown),
but it turns out that, due to the rather large overlap $\langle f_k|f_n\rangle$ of the different Lanczos vectors, the current Lanczos vector will be changed significantly by this procedure; which results in an even worse representation of the Hamiltonian. Alternatively speaking, the fact that re-orthogonalization involves sums implies again compression, which will make minor new approximations only if the re-orthogonalization is mild. Therefore we tried to devise a re-orthogonalization scheme which does not change the Lanczos vectors explicitly. 

To this end we rewrite the re-orthogonalization Eq.~\eqref{reortho} in the form
\begin{align}
|\psi_n\rangle = \sum^{n}_i S_{in} |f_i\rangle \label{reorthoS}\;,
\end{align}
where we have introduced the re-orthogonalization matrix $S_{in}$. It can be shown (see Appendix  \ref{App1} ) that 
$S_{in}$ can be calculated from the overlap elements\footnote{$W_{ij}$ is called the Gramian matrix. It gives a measure for the linear dependence of the Lanczos vectors. As long as $\det W_{ij}>0$ we still add linear independent vectors to the subspace. } $W_{ij}=\langle f_i|f_j\rangle$ by a recursion relation. The matrix elements of the effective Hamiltonian then
become
\begin{align}
\langle \psi_n|H|\psi_m\rangle = \sum^{n,m}_{i,j} S^\ast_{in}S_{jm} \langle f_i|H|f_j\rangle \, . \label{effReortho}
\end{align}
By construction the vectors $|\psi_i\rangle$  define an orthonormal basis system, i.e.,\ on the left hand side we have ensured 
that the effective Hamiltonian  is connected by a unitary transformation to the right hand
side. In particular, we do not have to change the calculated Lanczos 
vectors any more. The price that we are paying is that we need all overlap matrix elements 
$W_{ij}$ and matrix elements $\langle f_i|H|f_j\rangle$,\footnote{The calculation of the overlap elements and the matrix elements can be done in parallel. We want to remark that with the adaptive DMRG method we cannot calculate these overlap elements as we do not have all the Lanczos vectors in the same basis.} i.e.,\ we have to store all Lanczos vectors generated during 
the calculation. 
Finally, the effective Hamiltonian given by the matrix elements $\langle \psi_n|H|\psi_m\rangle$ is not tridiagonal anymore and we have to apply a full diagonalization to get the approximate 
excitation energies (The dimension of the effective Hamiltonian is given by the number of calculated Lanczos vectors and therefore the complete diagonalization is not computationally expensive at all;so this additional cost is minor).  

\begin{figure}[t!]
\centering
\includegraphics[width=0.98\columnwidth]{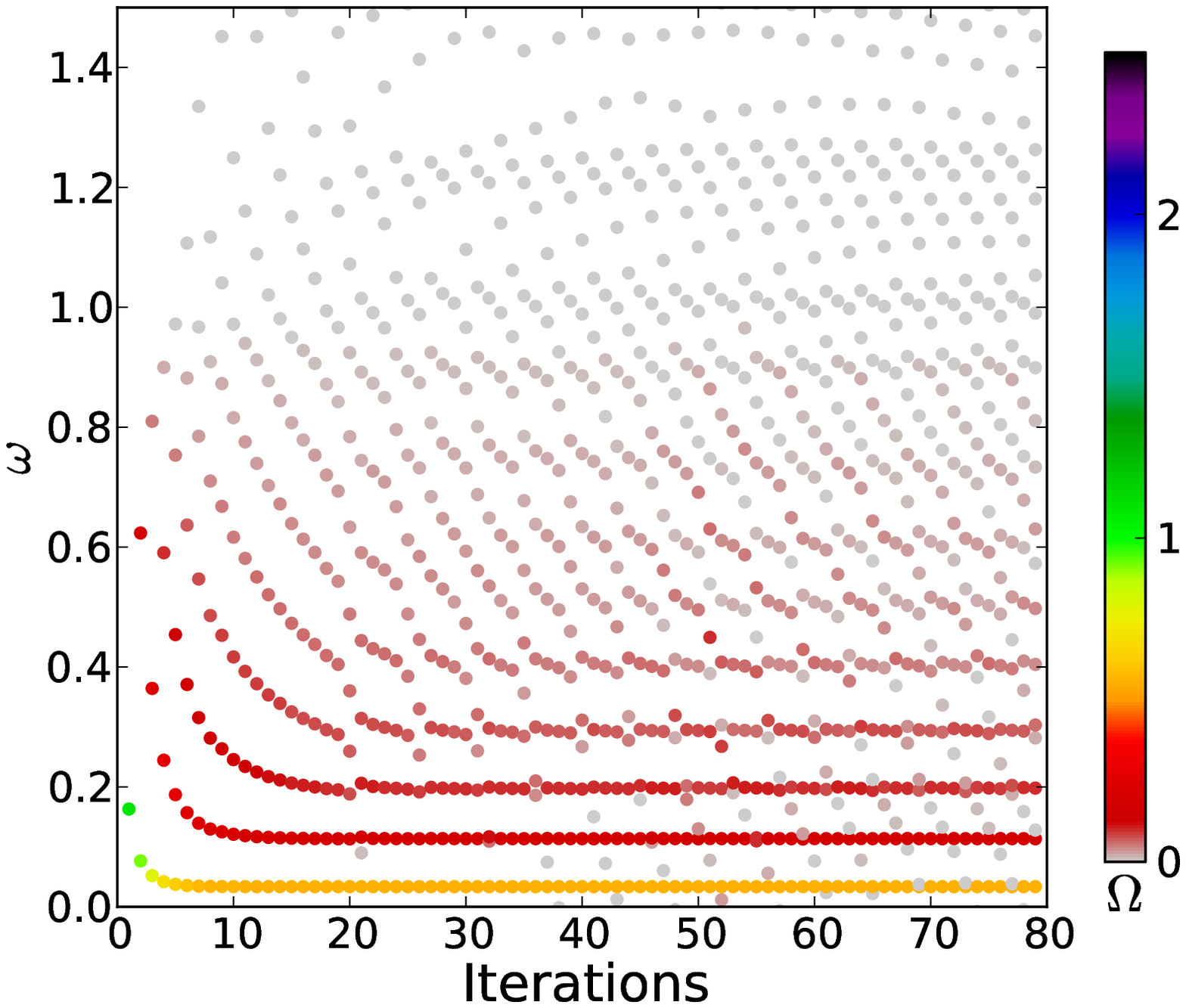}
\includegraphics[width=0.98\columnwidth]{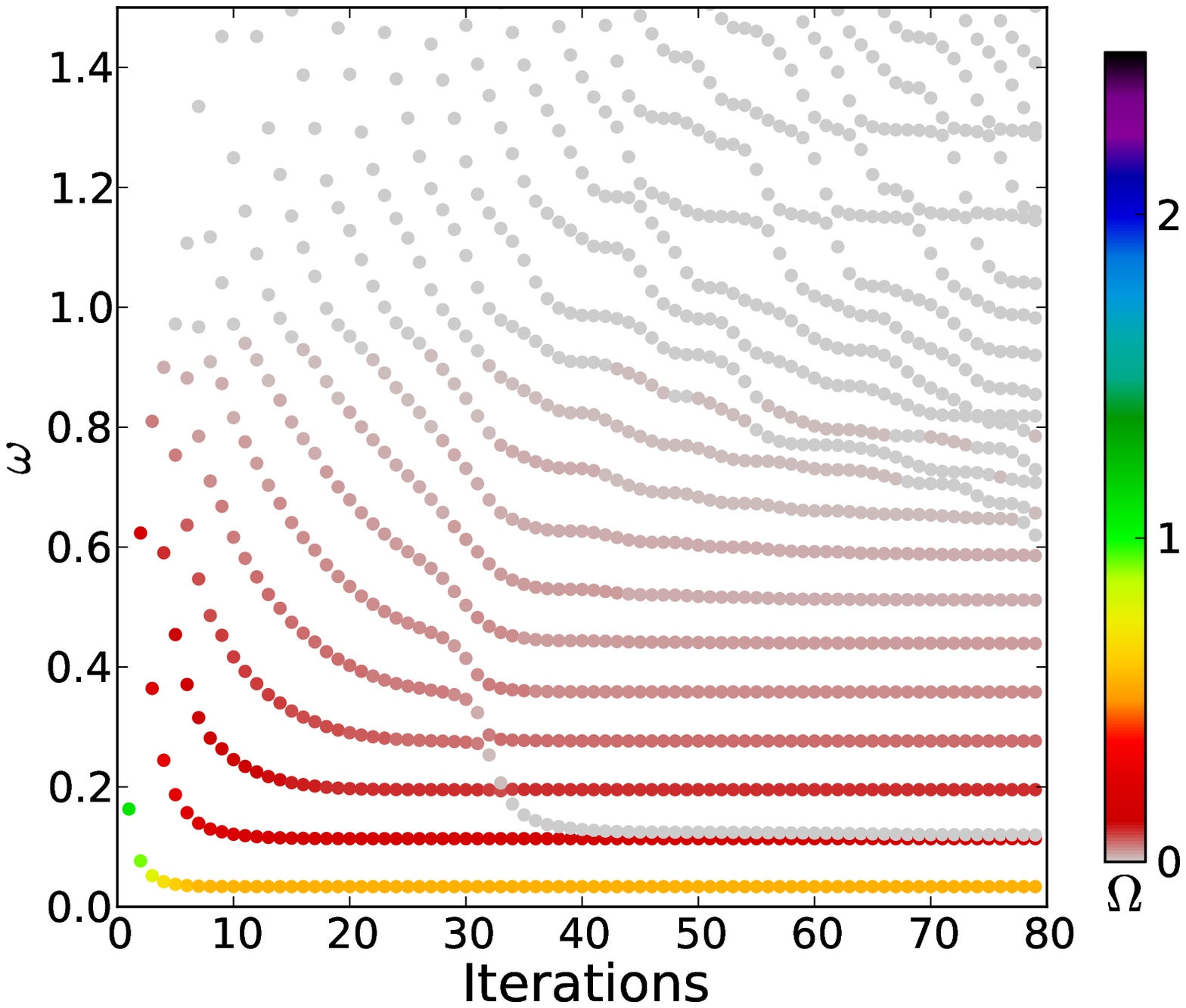}
\caption{\label{fig1} (Color online) Comparison of the convergen\label{key}ce of the 
excitation energies and spectral weights for pseudomomentum $k=\pi$ as function of Lanzcos iterations for the spin-$\frac12$ Heisenberg chain.
The upper panel shows the behavior for the scheme without any reorthogonalization, the lower 
panel the new scheme. Calculations were done for a chain of length $L=32$ (obc), maximal MPS matrix dimension $M=512$.}
\end{figure}  
Since the vectors $|\psi_i\rangle$ are orthonormal, the spectral weight $\Omega_k$ belonging to the 
excitation energy $\epsilon_k$ can now be calculated very easily from\footnote{This approach is
thus, besides being numerically more stable, also much simpler as compared to the adaptive Lanczos 
method,\cite{Dargel2011} where we had to evaluate an integral in the complex plane. }
\begin{align}
\Omega_k/\langle 0| \hat{A}^\dagger \hat{A} | 0 \rangle=|\langle E_k|A|0 \rangle|^2 = |\sum_i c_{ki} \langle \psi_i|\psi_0 \rangle|^2 = |c_{k0}|^2 \notag\;,
\end{align}
where we have used that the exact diagonalization of the effective Hamiltonian yields the (approximate) 
eigenvectors $|E_k\rangle=\sum_i c_{ki} |\psi_i \rangle$. In Fig.\ \ref{fig1} a comparison of the convergence of 
the spectral weights of the MPS Lanczos method with (lower panel) and without (upper panel) re-orthogonalization is shown. 
One can clearly see that with a maximal MPS matrix dimension $M=512$ only the first three spectral poles converge without the re-orthogonalization. With our reorthogonalization scheme one obtains
a much better and more stable convergence, which allows to extract already the first $8-9$ poles even for this 
small maximal MPS matrix dimension.  Note however, that one of course does not get rid of all the ghosts. 

The obvious question within our algorithm thus is when to stop the Lanczos iterations and how to distinguish real
excitation energies from ghosts. The usual criterion for the quality of an (approximate) pole position
$\omega_k$ is the residual. The residual vector is defined via
\begin{align}
 |r_k\rangle=H |E_k\rangle -\omega_k |E_k\rangle\notag  
 \end{align}
 and the residual by 
\begin{align}
r_k = \langle r_k| r_k \rangle = \langle E_k | (H-\omega_k)^2 | E_k\rangle \notag\;.
\end{align}
The residual can be viewed as a measure, how well the pole $\omega_k$ and (normalized) vector $|E_k\rangle$ approximate an eigenvalue and eigenvector of $H$.
We can again express this residual without knowing the eigenvectors explicitly, for the additional price
that we have to measure all the matrix elements for $H^2$. Then the quantities entering the residual can be 
calculated as
\begin{align}
\langle E_k | H | E_k\rangle &= \sum_{ij}  c_{ki}c_{kj}\sum^{i,j}_{p,q}  S_{pi} S_{qj} \langle \notag f_p|H|f_q\rangle \, , \notag\\
\langle E_k | H^2 | E_k\rangle &=  \sum_{ij}  c_{ki}c_{kj} \langle \psi_i | H^2 | \psi_j \rangle \notag\\
&=\sum_{ij}  c_{ki}c_{kj}\sum^{i,j}_{p,q}  S_{pi} S_{qj} \langle f_p |H^2| f_q\rangle \, .\notag
\end{align}
With the help of the residuals we can define the following algorithm to calculate the position and weight
of the poles: 
\begin{enumerate}
\item Remove all poles with a spectral weight below $\Omega_{cut}$ and a residual that is larger than 
$r_{cut}$, because they very likely are ghosts. 
\item Follow the convergence of each of the remaining poles and take those with the smallest residual.
\end{enumerate} 

 \begin{figure}[t!]
 \centering
 \includegraphics[width=0.95\columnwidth]{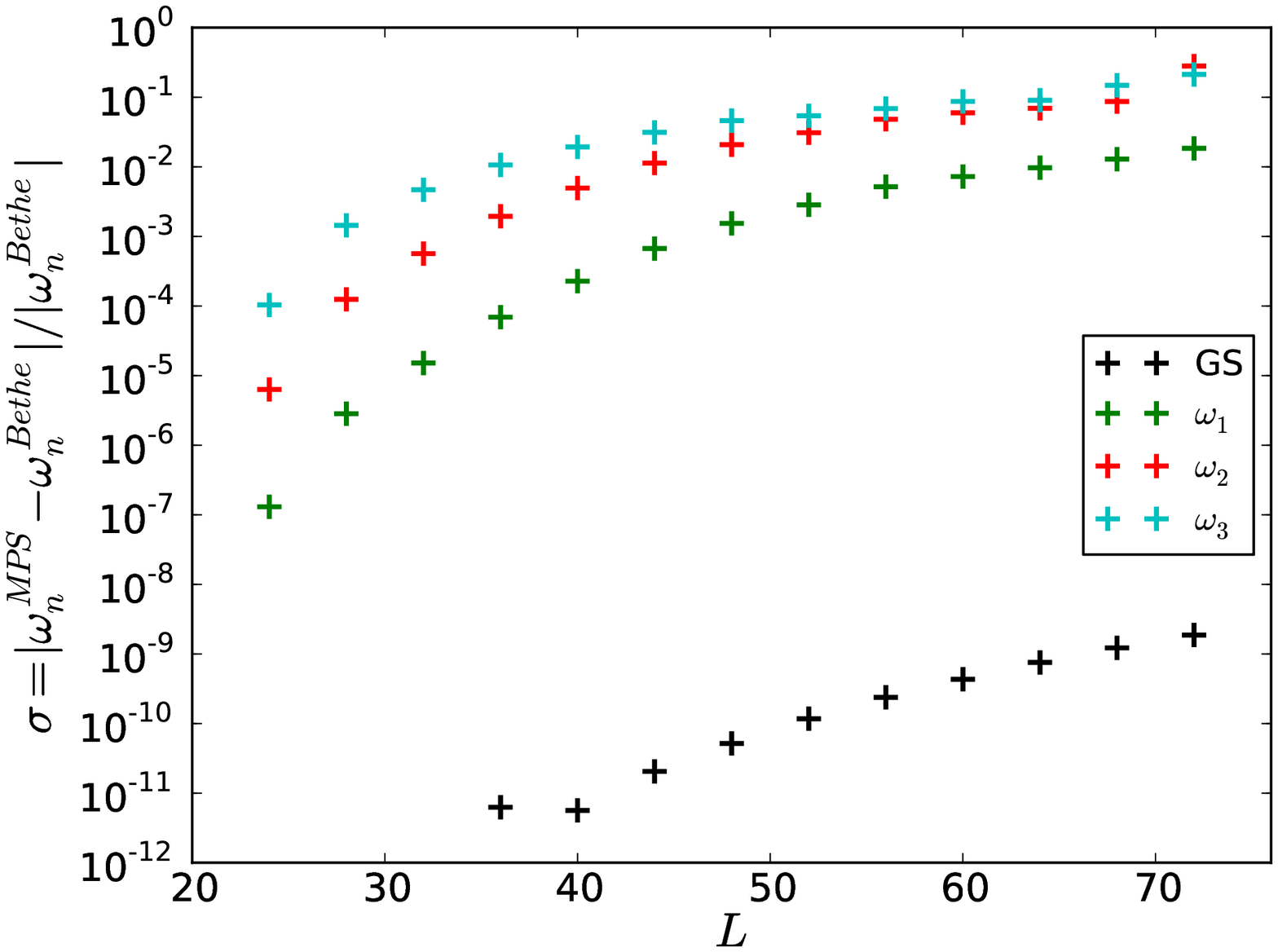}
 \includegraphics[width=0.95\columnwidth]{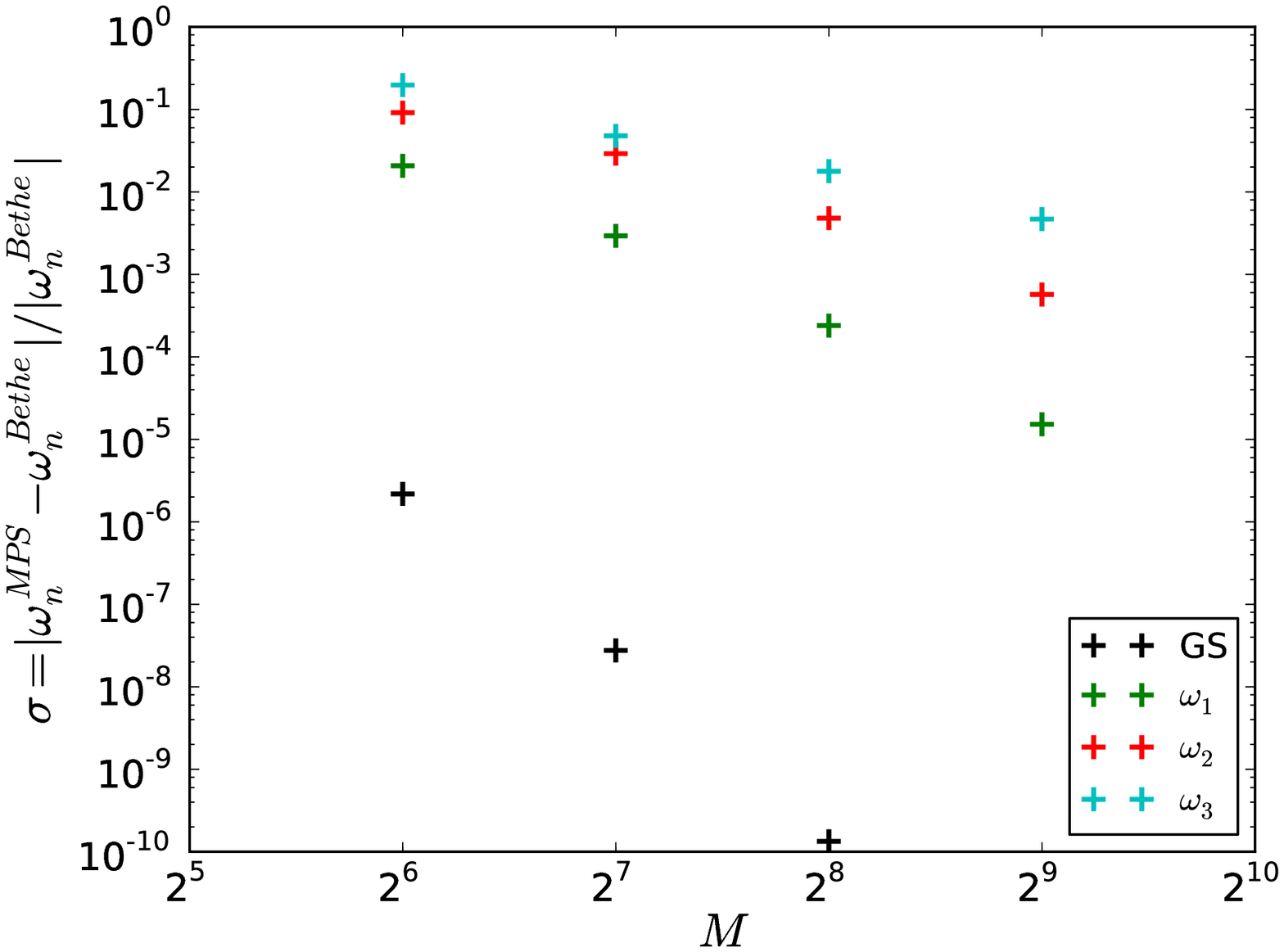}
 \caption{\label{fig2} (Color online) The relative error of the first three spectral poles ($k=\pi$,pbc) calculated with the MPS Lanczos method in comparison to the two-spinon excitations from the Bethe Ansatz for different system sizes (upper panel) and for $L=32$ for different maximal MPS matrix dimension $M$ (lower panel). The groundstate (GS) error is also shown. }
 \end{figure}  
In Fig.\ \ref{fig2} the relative error in the position of the first three excitations for $k=\pi$ and the groundstate energy calculated for periodic boundary conditions is shown. We compare
our numerical results to exact ones based on the Bethe ansatz. The weight cutoff in our algorithm was chosen as
$\Omega_{cut}=10^{-3}$ in all cases. The residual cutoff had to be increased with system size from
$r_{cut}=0.1$ for $L=24$ to $r_{cut}=0.5$ for $L=72$ to be able to reliably extract the spectral weights. As expected one finds  
that the error in the groundstate energy is several orders of magnitude smaller than the error for the
excitations. In the upper panel  of Fig.\ \ref{fig2} the system size $L$ is varied for a constant maximal MPS matrix dimension $M$, while in the lower panel we keep $L=32$ fixed and increase the maximal MPS matrix dimension $M$.
The monotonic decrease of the error shows that it originates chiefly in the approximation of the 
Lanczos states by MPS.

In the MPS-based Lanczos method we have now obtained accurate poles and spectral weights for a given finite chain length. In order to extrapolate these to the thermodynamic limit we will use the rescaling scheme described in section \ref{subsec:basicLanczos}.

\section{Results}
\label{sec:results}

\subsubsection{$k=\pi$}

\begin{figure}[tb]
\centering
\includegraphics[width=0.98\columnwidth]{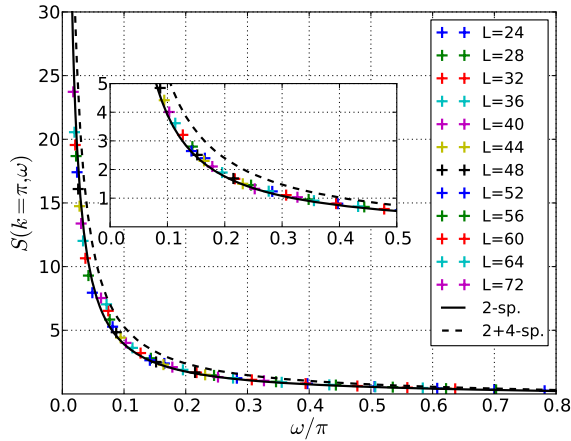}
\includegraphics[width=0.98\columnwidth]{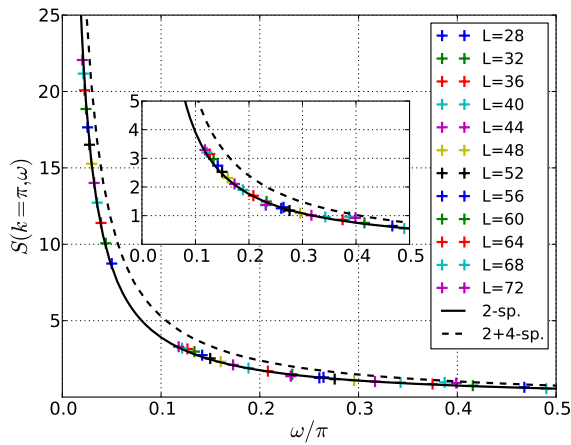}
\caption{\label{fig3} (Color online) Dynamical correlation function for $k=\pi$ for open (upper panel) and periodic boundary (lower panel) conditions. The spectral weights/poles are chosen by the minimal residual scheme.}
\end{figure}

We have tested our Lanczos algorithm for $S(k=\pi,\omega)$ for both open and periodic boundary conditions.
The results are collected in Fig.~~\ref{fig3}.  The energy cutoff for the poles was chosen as 
$\Omega_{cut}=10^{-3}$, and the residual cutoff had to be increased with system size to maximal 
$r_{cut}=0.5$ to be able to extract the spectral weights.
One can clearly see that  the discrete spectral weights, rescaled with the scheme discussed in section
\ref{subsec:basicLanczos}, nicely collapse onto a single curve, which is in good agreement with the results from the 
2-spinon contributions of the Bethe ansatz in the thermodynamic limit,\footnote{The Bethe ansatz data is provided by J.S. Caux, see Ref.~\onlinecite{Caux2006}.} independent of the chosen boundary conditions. Finite size effects are not very pronounced for $k=\pi$, and all (significant) spectral weight lies
within the two-spinon bounds (see Eq.~\eqref{bounds}) $\omega_1(k=\pi)=0$ to $\omega_2(k=\pi)=\pi$.
The position of the second spectral pole differs substantially between open and periodic boundary conditions. 
Therefore for periodic boundary conditions we would need much longer chains to fill the gap between the first 
and second spectral poles. The position of the first spectral pole moves very slowly with system size to the origin. Therefore, in order to resolve the divergence for $\omega\rightarrow 0$ we would have to go to much larger system sizes. The eigenstates that lead to non-vanishing weights for the dynamical spin structure factor belong all to the $S=1$ quantum sector.\cite{Muller1981} For $k=\pi$ it is the eigenvector with the lowest energy in that subspace that gives the first spectral weight/pole. Therefore an alternative way to check the behavior for $\omega \rightarrow 0$  for $k=\pi$ is to do standard groundstate DMRG calculations in that subspace (not shown). But in order to use our rescaling scheme we need also the subsequent poles which are not as easy to obtain as long as translation symmetry is not implemented.\footnote{The next eigenvectors that can be obtained by a multi-target approach have no weight for $S(k=\pi,\omega)$ as they have a different momentum. Therefore in order to use the method efficiently we have  to search directly in that momentum subspace. That means we have use tranlational symmetry and implement momentum as a good quantumnumber.}

 It turns out that the Lanczos method does not capture the four-spinon weights (and higher spinon weights). In principle the Lanczos algorithm will also converge towards any multi-spinon state. But for this model the four-spinon and higher-spinon states show a rather strong finite size scaling,\footnote{ For example, the first four-spinon pole lies at $\omega=2.41$ for $L=24$, see Ref.~\onlinecite{Faribault2009} and private communication with J.S. Caux.} which means that  four-spinon states will appear in the low energy sector only for much longer chains 
 than those considered here. Furthermore, these states will have a small weight in comparison to 
the two-spinon states and therefore they will be hard to distinguish from ghost states or numerical noise.

\begin{figure}[tb]
\centering
\includegraphics[width=0.98\columnwidth]{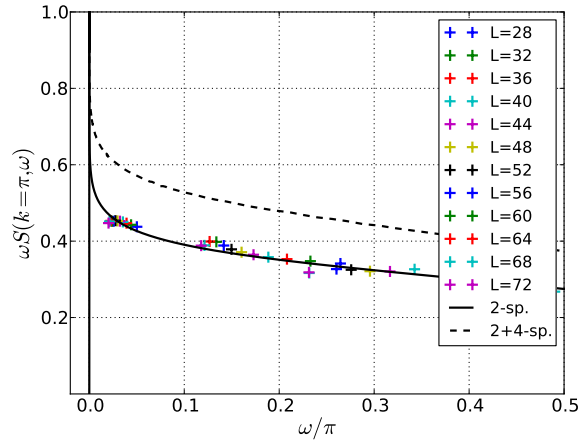}
\caption{\label{fig4} (Color online)  Rescaled dynamical structure factor  $\omega S(k,\omega)$ for $k=\pi$ and periodic boundary conditions. This figure is based on the same data as in Fig. \ref{fig3}.}
\end{figure}

\begin{figure}[tb]
\centering
\includegraphics[width=0.98\columnwidth]{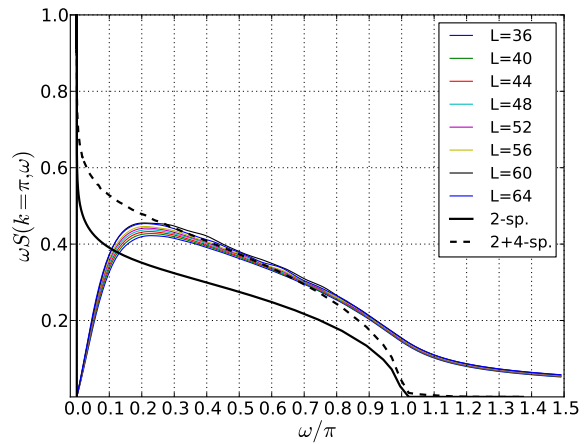}
\caption{\label{fig5} (Color online)   Rescaled dynamical structure factor  $\omega S(k,\omega)$ for $k=\pi$ and open boundary conditions with a Lorentzian broadening of $\eta=0.1$.}
\end{figure}

\begin{figure}[tb]
\centering
\includegraphics[width=0.75\columnwidth,angle=270]{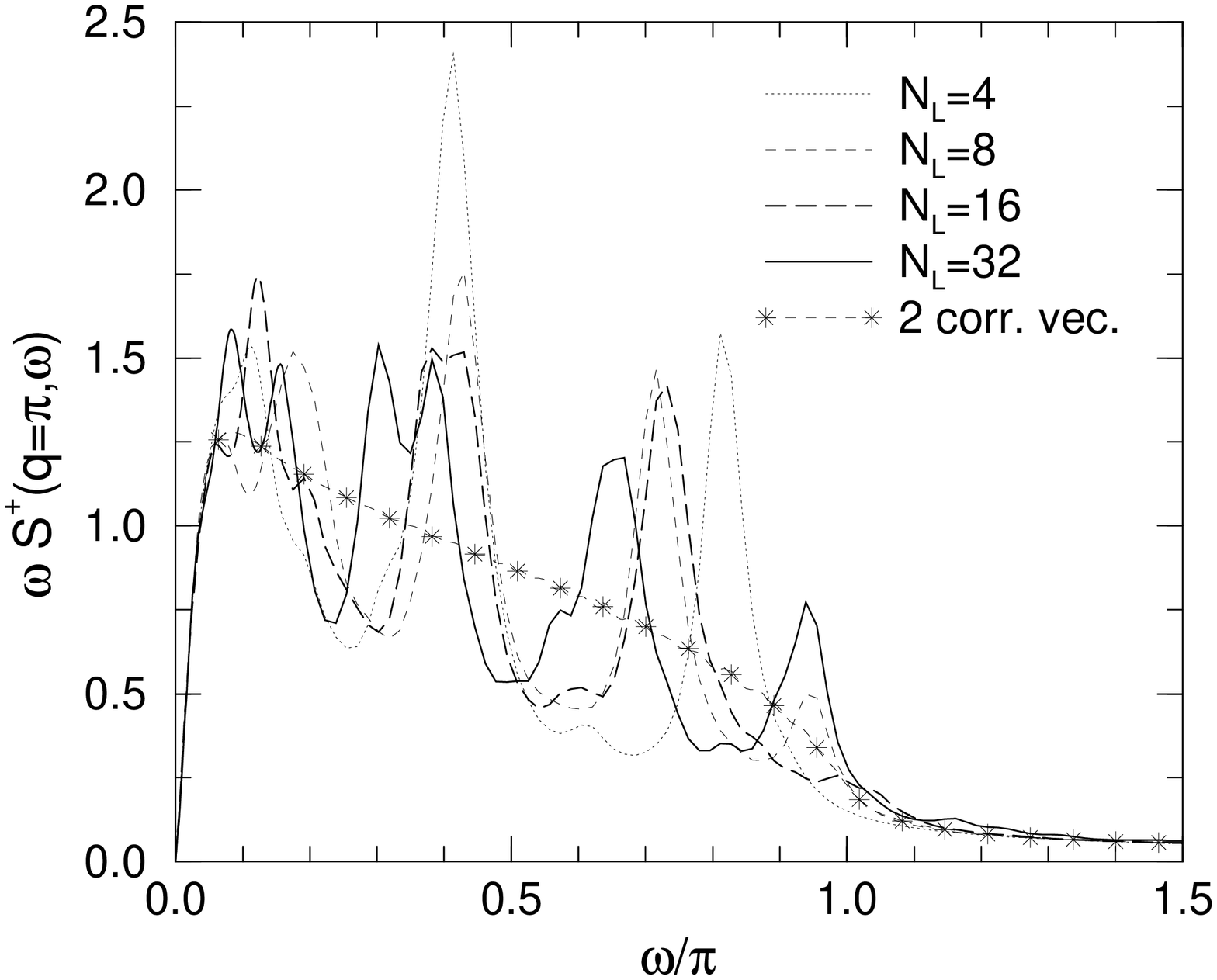}
\caption{\label{fig6} Dynamical structure factor  $\omega S^{+}(\omega,k)$ for $k=\pi$ and open boundary conditions with a Lorentzian broadening of $\eta=0.1$, obtained in a standard Lanczos approach (with various numbers of target states $N_L$) and in the correction vector approach; reproduced from Fig.~8 of Ref.~\onlinecite{Kuhner1999}. Note that due to different schemes of dealing with the open boundary conditions results relate to those in Fig.~\ref{fig5} only up to a proportionality constant.}
\end{figure}

As mentioned earlier, the structure factor $S(k=\pi,\omega)$ shows a logarithmic correction to the
leading divergence as $\omega\to0$ (see Eq.\ \eqref{div}).
In order to resolve such an additional feature it is instructive to plot $\omega S(k,\omega,k)$, which is
done in Fig.\ \ref{fig4} for the periodic boundary conditions.
Apparently, the data nicely fall onto the line from the Bethe ansatz, i.e.\ are in agreement with the predicted
logarithmic divergence of this quantity. In order to make contact to earlier Lanczos methods, in Fig.\ \ref{fig5}, $\omega S(k=\pi,\omega)$ is plotted again, but for open boundary conditions and with an additional Lorentzian broadening of $\eta=0.1$. For this figure we do not have to use any residual cutoffs or weight cutoffs as for the extraction of the single weights. The broadening completely smears out the logarithmic divergence, but this is not specific to our method. In Figs.\ 8 and 14 of  Ref.\ \onlinecite{Kuhner1999} the same quantity  up to a proportionality constant is discussed by K\"uhner and White (their Fig.\ 8 is reproduced here for comparison, Fig.\ \ref{fig6}). In their calculation, they used the standard DMRG Lanczos (broadened) and the (numerically much more expensive) correction vector method.   Although different schemes for dealing with the open boundary conditions preclude a quantitative comparison, their results clearly show that the standard Lanczos DMRG is not capable of producing a smooth, converged 
curve consistent with the predictions from Bethe ansatz, but shows substantial artificial structure. This comparison clearly shows that we have reached
a definite improvement over the standard Lanczos method. We want to emphasize that the inherent broadening of the correction vector completely annihilates any signature of the divergence for $\omega \rightarrow 0$ (c.f. Fig.\ 14, Ref.~\onlinecite{Kuhner1999}).
Note, however, that in order to unambiguoulsy resolve or even 
predict such a  logarithmic divergence we would need to study system sizes well beyond anything possible 
presently.

\subsubsection{$k=\frac{\pi}{2}$}
\begin{figure}[tb]
\centering
\includegraphics[width=0.98\columnwidth]{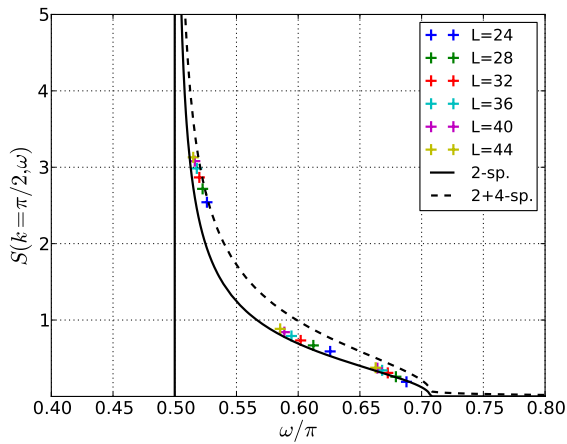}
\caption{\label{fig7} (Color online)  Dynamical correlation function for $k=\pi/2$ for periodic boundary conditions and $M=512$. The spectral weights/poles are chosen by the minimal residual scheme.     }
\end{figure}
In Fig.\ \ref{fig7} we show $S(k=\frac{\pi}{2},\omega)$ (periodic boundary condictions) calculated with our
method compared to the 2-spinon contribution obtained from Bethe ansatz.\cite{Caux2006} The spectral signature starts now
in the middle of the spectrum and therefore the extraction of poles and weights is much harder. We were
able to obtain reasonable data up to a length $L=44$ and $M=512$. The low-lying excitations do not collapse 
as nicely onto the 2-spinon curve as for $k=\pi$. One reason may lie in the problematic definition of the first 
interval in the rescaling scheme, which explicitly depends on the value of the lower bound, $\omega_1$.
It is evident that this value will also be subject to more or less strong finite size effects. These finite size effects are obviously even worse for open boundary conditions even though we can go to larger systems.

\section{Discussion}
\label{sec:discussion}

As we have seen, the adaptation of the Lanczos algorithm for dynamical correlation functions to the framework of matrix product states produces much more accurate results than previous implementations in the DMRG framework and is capable of handling the broadly distributed spectral weight of the $S=\frac{1}{2}$ Heisenberg antiferromagnet very well, which had been a major stumbling block for the original DMRG-based formulation. In more technical terms, the new algorithm 
offers three main advantages over previous Lanczos implementations in the DMRG:
\\
(i) We have shown that the Lanczos algorithm in MPS formulation is capable of calculating a number of discrete spectral weights/poles for spin chains that are longer than those that can be evaluated with exact diagonalization and the standard Lanczos algorithm. The algorithm can be easily implemented in any MPS algorithm as it just uses simple MPS algebra routines. It improves over the Lanczos implementations in classical DMRG and the adaptive Lanczos DMRG algorithm by avoiding multi-targeting and the inexact Hamiltonian representation. 
\\
(ii)
Furthermore the possibility of handling the Lanczos states individually by a single MPS allows for re-orthogonalization procedures that are not possible with the adaptive Lanczos method. This re-orthogonalization significantly improves the convergence of the spectral poles and makes the extraction of single weights much easier.
\\
 (iii) As the number of discrete spectral poles is very limited (in the spin-$\frac{1}{2}$ Heisenberg chain) it is nearly impossible to get valuable information for the behavior in the thermodynamic limit from one single chain in this length scale. One should emphasize that all the information about the spectral function is just given by these discrete spectral weights/poles and therefore any method for the
 DMRG/MPS that gives a continuous function (e.g. correction vector, expansion in Chebyshev polynomials, Lanczos plus continuous fraction) just interpolates between these points depending on the broadening scheme of the used method (See section \ref{subsec:basicLanczos}). The used rescaling scheme based on the explicit extraction of discrete spectral weights and poles gives a very controlled access to the problematic limit $\eta\to0, L\to \infty$ and by this gives more transparent results without broadening.\\
 
We expect that the main competitor of this approach will be the other new numerical low-cost method provided by expansion in Chebychev polynomials. Compared to each other, both have advantages and disadvantages, such that no conclusive statement can be made at the moment, also in view of the lack of applications so far. Both rely on operations $H|\psi\rangle$ carried out a similar number of times, but the Chebychev approach can handle longer chains ($\propto 300$ sites) because of a surprising low maximal MPS matrix dimension ($M \sim 32-64$); this is arguably due to an inherent entanglement-reducing energy-truncation scheme. However, comparing directly the evaluation of discrete spectral weights/poles, one has to perform many iterations (500-1000) within the Chebyshev expansion. Furthermore, one has to fit the broadened spectral weights to get the discrete poles/weights including a fit error. This becomes even more challenging for dense spectral poles, where the broadened peaks overlap. Generally, broadening is an inherent part of the Chebychev approach. It also has more tuning parameters that have to be optimized for the method to work properly.   \\

Some algorithms for time evolution of quantum states also use a Krylov space representation in  Matrix Product States. \cite{Schmitteckert2004,Noack2005,Manmana2005,Garcia2006,Kleine2008,Flesch2008} Here our proposed re-orthogonalization may also be useful.
\\ 

Recently, two algorithms for calculating spectra of translationally invariant Hamiltonians with (unitary) MPS working directly in the thermodynamic limit\cite{Haegeman2011}/ with periodic boundary conditions\cite{Pirvu2012} were suggested that can approximate many branches of the dispersion relation with high precision. It would be really interesting to see whether these algorithms can also calculate dynamic correlation functions with that precision and how they compare to the Lanczos expansion. \\

\begin{acknowledgments}
We would like to thank Reinhard Noack and Didier Poilblanc for fruitful discussions. Furthermore we would like to thank Jean-Sebastien Caux for the thermodynamic Bethe Ansatz data and for information on single four-spinon weights and Steve White for permission to reproduce the present Fig. \ref{fig6} from Ref.~\onlinecite{Kuhner1999}, Fig.~8. 
We acknowledge computer support by the Gesellschaft für wissenschaftliche Datenverarbeitung Göttingen (GWDG) and we would like to thank the Deutsche Forschungsgemeinschaft for financial support via the collaborative research center SFB 602 (TP A3) and a Heisenberg fellowship (Grant No. 2325/4-2, A. Honecker)\\

\end{acknowledgments}

\appendix
\section{Recursive formula for reorthogonalization} \label{App1}
We want to express the reorthogonalization equation
\begin{align}
|\tilde \psi_n\rangle &=|f_n\rangle-\sum^{n-1}_i \langle f_n|\psi_i\rangle |\psi_i\rangle, \quad 
|\psi_n\rangle &= \frac{1}{N_n}|\tilde \psi_n\rangle
\label{reorthoApp}
\end{align}
 by a matrix $S$ with $|\psi_n\rangle = \sum^{n}_i S_{in} |f_i\rangle$ (with $N_n=\sqrt{\langle\tilde\psi_n|\tilde\psi_n\rangle}$). This leads to:
\begin{align}
|\tilde \psi_n\rangle &= |f_n\rangle-\sum^{n-1}_i  \langle f_n|\psi_i\rangle |\psi_i\rangle\notag\\
&= |f_n\rangle-\sum^{n-1}_i \langle f_n|\left(\sum^{i}_{k} S_{ki} |f_{k}\rangle\right) \left(\sum^{i}_{k'} S_{k'i} |f_{k'}\rangle\right)\notag \\
&= |f_n\rangle-\sum^{n-1}_i \left(\sum^{i}_{k} S_{ki} W_{nk}\right) \left(\sum^{i}_{k'} S_{k'i} |f_{k'}\rangle\right)\notag\\
&= |f_n\rangle-\sum^{n-1}_i \sum^{i}_{k'} k_{in} S_{k'i} |f_{k'}\rangle\notag,\\
k_{in}&=\sum^{i}_{k} S_{ki} W_{nk},\quad W_{ij}=\langle f_i|f_j\rangle\,. \notag
\end{align}
Now one can deduce a recursion formula for the matrix elements of $S$:
\begin{align}
\tilde S_{pn}&=\begin{cases}
-\sum^{n-1}_{q=p}k_{qn}S_{pq},  & \text{if } p<n\\
1, & \text{if }p=n
\end{cases}
\end{align}
with
\begin{align}
S_{pn}&=\frac{1}{N_n} \tilde S_{pn},\quad S_{00}=\frac{1}{N_0}\notag\\
N_n&=\sqrt{\sum_{q,p=0}^n \tilde S_{pn} \tilde S_{qn} W_{pq}},\quad N_0=\sqrt{W_{00}}\,.\notag
\end{align}

\bibliography{Literatur}

\end{document}